\documentclass{vgtc}                          % final (conference style)
\graphicspath{{figures/}{pictures/}{images/}{./}} % where to search for the images
\usepackage{times}                     % we use Times as the main font
\usepackage{tabu}                      % only used for the table example
\usepackage{booktabs}                  % only used for the table example
\usepackage{lipsum}                    % used to generate placeholder text
\usepackage{mwe}                       % used to generate placeholder figures

\usepackage{enumitem}

\usepackage{balance}

\UseRawInputEncoding

\usepackage{mathptmx}                  % use matching math font

\newcommand{\revision}[1]{\textcolor{black}{#1}}

\onlineid{0}

\vgtccategory{Research}

\vgtcinsertpkg

\title{VR as a ``Drop-In'' Well-being Tool for Knowledge Workers}

\author{Sophia Ppali\thanks{e-mail: s.ppali@cyens.org.cy} %
\and Haris Psallidopoulos\thanks{e-mail: h.psallidopoulos@cyens.org.cy} %
\and Marios Constantinides\thanks{e-mail: marios.constantinides@cyens.org.cy} %
\and Fotis Liarokapis\thanks{e-mail: f.liarokapis@cyens.org.cy}}
\affiliation{\scriptsize CYENS Centre of Excellence}

\teaser{
  \centering 
  \includegraphics[width=\textwidth]{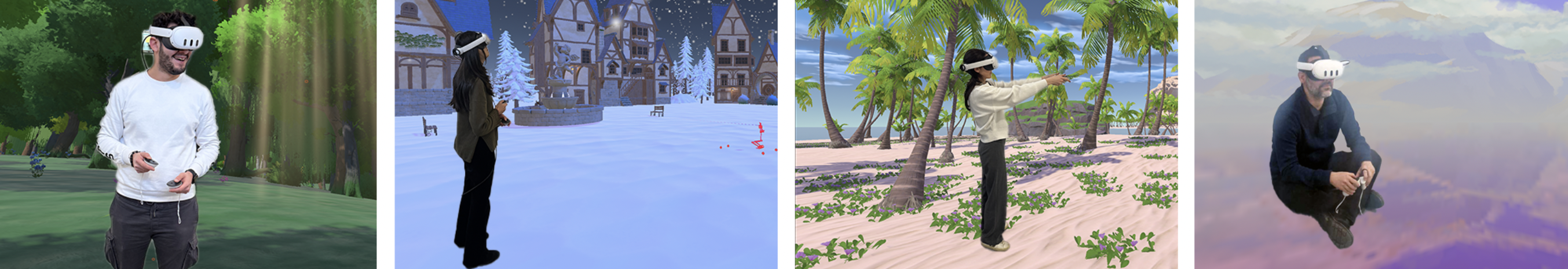}
  \caption{Tranquil Loom is a VR app designed for short and self-directed breaks at work. It offers knowledge workers a choice of calming environments (forest, snow, beach, abstract) and activities (short stretching session, guided meditation, and open-ended exploration) to support mental, physical, and cognitive well-being.}
  \label{fig:teaser}
}

\abstract{Virtual Reality (VR) is increasingly being used to support workplace well-being, but many interventions focus narrowly on a single activity or goal. Our work explores how VR can meet the diverse physical and mental needs of knowledge workers. We developed Tranquil Loom, a VR app offering stretching, guided meditation, and open exploration across four environments. The app includes an AI assistant that suggests activities based on users’ emotional states. We conducted a two-phase mixed-methods study: (1) interviews with 10 knowledge workers to guide the app’s design, and (2) deployment with 35 participants gathering usage data, well-being measures, and interviews. Results showed increases in mindfulness and reductions in anxiety. Participants enjoyed both structured and open-ended activities, often using the app playfully. While AI suggestions were used infrequently, they prompted ideas for future personalization. Overall, participants viewed VR as a flexible, ``drop-in'' tool, highlighting its value for situational rather than prescriptive well-being support.} %\todo{1. check throughout the paper for the word/phrase "on demand" if so, remove it. 2. normalize the use of "app", "tool", "application", "system". 3. write everything in American English}

\keywords{virtual reality, well-being, future of work, knowledge workers, human-centered design}

\begin{document}


% \firstsection{Introduction}
\firstsection{Introduction}

\maketitle

Virtual Reality (VR) is increasingly being explored as a tool for supporting well-being across a range of contexts, including the workplace \cite{Riches2023Virtual}. Its immersive nature allows users to temporarily step away from their immediate environment~\cite{slater_enhancing_2016} by offering a form of instant detachment, especially for those in mentally intensive and sedentary roles. Prior research has demonstrated the potential of VR in promoting mental restoration ~\cite{ma_effectiveness_2023}, reducing stress ~\cite{xu_effectiveness_2024}, and supporting light physical activity ~\cite{kim_effect_2021, yoo_embedding_2020}.  Additionally, there is empirical evidence that VR apps such as nature-based relaxation environments or guided meditations can promote mental clarity~\cite{ladakis2024virtual}.  As organizations pay greater attention to employee well-being, immersive technologies have emerged as potential tools for helping people disconnect temporarily from busy or screen-heavy environments, re-center during stressful moments, or re-energize between tasks~\cite{Riches2023Virtual, wagener_role_2021}. In workspace settings, VR can offer an immediate shift in environment without requiring employees to leave their workspace, which may be particularly useful for those with limited time or flexibility during the workday.

This is especially relevant for knowledge workers working in office settings. Typically, knowledge workers deal with processing, analyzing, and managing information~\cite{Surawski2019Who}. As central drivers of productivity in many sectors~\cite{Greene2011Space}, they frequently navigate cognitive overload, long periods of physical inactivity, and blurred boundaries between work and life~\cite{wepfer_work-life_2018, soto_observing_2021}. Such demands can affect mental focus, physical comfort, and overall well-being~\cite{wepfer_work-life_2018}. While organizations offer interventions including flexible schedules or wellness programs~\cite{teevan_new_2021, chow_feeling_2024}, these do not always align with how people actually manage their time and energy during the day. At the same time, many knowledge workers are comfortable experimenting with digital tools to support their productivity and focus~\cite{guillou_is_2020}. For example, a recent report by Microsoft and LinkedIn found that 75\% of knowledge workers already use AI in their roles, with most reporting increased efficiency and creativity~\cite{MicrosoftLinkedIn2024}.  

Openness to innovation presents an opportunity for wellness interventions through technology. One such technology is VR. Yet, current VR tools for the workplace often adopt a solution-focused mindset, offering single-purpose experiences such as guided meditation or virtual nature walks \cite{adhyaru2022virtual, Riches2023Virtual}. They rarely account for the complexity of real-world work where needs vary widely across individuals and time. Broader literature on VR for well-being highlights the need for more flexible and personalized approaches \cite{wagener_role_2021}. Still, little research has explored what this might mean in the specific context of everyday knowledge work. Different workers may need different forms of support at different times, whether it is movement to relieve tension, a quiet space to reset, or playful exploration to re-energize. A tool that assumes a single definition of well-being or imposes a fixed routine is unlikely to support a wide range of users in meaningful ways. Moreover, the question of how such tools could be integrated into the rhythms and constraints of working life remains largely underexplored~\cite{Riches2023Virtual}.

To address this gap, our work investigates how VR might support the well-being of knowledge workers in ways that are responsive to their everyday needs. Rather than developing new interaction modes or a novel technical app, our focus is on investigating how familiar well-being practices (i.e., stretching, meditation, and exploratory interaction) can be integrated into a single tool in ways that feel responsive to the pace and structure of everyday work. We position our work as an exploratory design study that surfaces the trade-offs and constraints of integrating diverse activity modes in VR, and reflect on how existing interventions can be adapted and situated in office settings. Rather than introducing new technical designs, we aim to surface lived user experiences and practical design implications for real-world workplace well-being.

\begin{figure*}[h!]
  \centering

  \includegraphics[width=.98\textwidth, height=3.9cm]{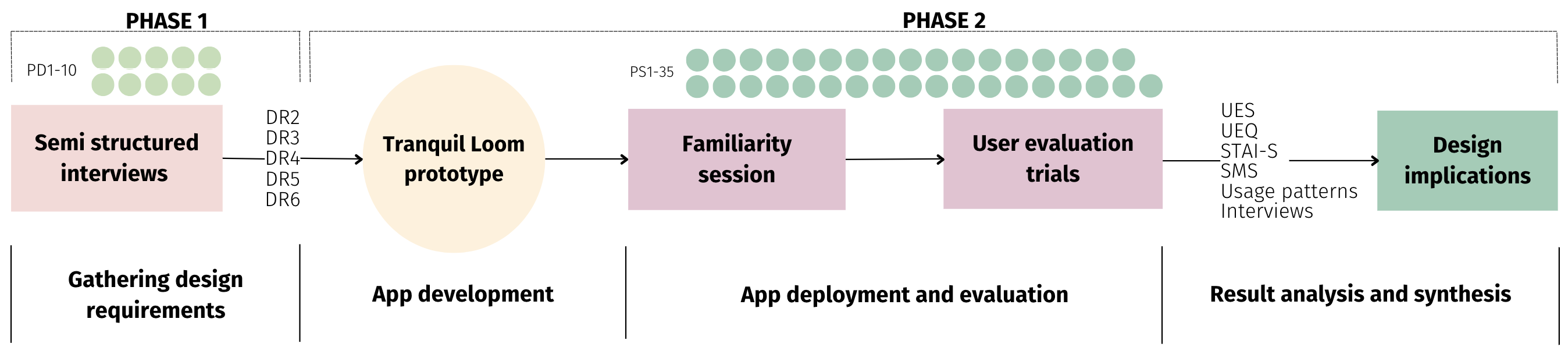} 

\caption{Two-phase study methodology. In Phase 1, we conducted semi-structured interviews with 10 knowledge workers to gather design requirements that informed the development of a VR well-being app \emph{Tranquil Loom}. In Phase 2, we deployed the app in a workplace setting with 35 participants and evaluated it through pre-post well-being measures, usage data, and follow-up interviews.}
\label{fig:methodology}

\end{figure*}

To achieve this, we carried out a two-phase mixed-methods study. In Phase 1, we conducted semi-structured interviews with 10 knowledge workers, during which we explored how participants manage their well-being during the day, their expectations and concerns around using VR in the workplace, and the types of experiences they believed would be helpful. Their insights informed the design of \textit{Tranquil Loom}, a VR app offering short guided stretching exercises, meditation, and open-ended exploration across four calming environments (Figure \ref{fig:teaser}). Tranquil Loom was designed to support different types of well-being needs and allow users to choose how to engage, depending on how they felt at the time, also supported with suggestions by an AI assistant. In Phase 2, we deployed the app in an office workplace setting with 35 participants. We combined usage data, pre- and post-intervention well-being measures, and follow-up interviews to examine how participants used the app, what they found helpful or unhelpful, and how VR might fit into the flow of a working day (Figure \ref{fig:methodology}). Our aim was to understand how VR might be designed and integrated in ways that reflect the challenges of everyday work. To achieve this goal, our work is guided by three research questions (RQs): 

\begin{enumerate}[label=\textbf{RQ\textsubscript{\arabic*}:}, align=left, leftmargin=2.2em, labelsep=0.5em, itemsep=0.8em]
    \vspace{-0.2cm} \item Which aspects of workplace well-being can VR address?
    \vspace{-0.2cm} \item How can VR experiences be tailored to reflect the varied personal and organizational needs of knowledge workers?
    \vspace{-0.2cm} \item Which design elements and functionalities are most effective for supporting knowledge workers' well-being within VR?
\end{enumerate}

In answering our RQs, we made three main contributions:

\begin{enumerate}
    \vspace{-0.2cm} \item Through semi-structured interviews with knowledge workers (\S\ref{sec:design}), we identified six design requirements for VR apps that account for the diverse demands of cognitive, emotional, and physical well-being in real-world work settings.

     \vspace{-0.2cm} \item With these requirements at hand, we designed and deployed a VR well-being app called \emph{Tranquil Loom} (\S\ref{sec:app}). The app featured three types of well-being practices (i.e., stretching, meditation, and exploration) across four environments. 
     
    \vspace{-0.2cm} \item A mixed-methods evaluation highlighting the design tensions and opportunities in VR-based workplace well-being (\S\ref{sec:evaluation}). We found that workers embraced VR as a situational ``drop-in'' tool rather than a scheduled routine and valued playful and self-directed engagement over AI-guided suggestions.

\end{enumerate}

With our work, we shift the focus from evaluating VR well-being apps as fixed interventions to designing for the complexity of real-world needs. We identified key design trade-offs such as between structure \emph{vs.} openness, doing \emph{vs.} being, and AI guidance \emph{vs.} autonomy (\S\ref{sec:discussion}), and discussed design implications for emotionally responsive and ethically grounded tools that support trust and spontaneous use in the workplace (\S\ref{sec:design-implications}).

\section{Related Work}
\label{sec:related-work}

We surveyed various lines of research that our work draws upon, and grouped them into two
areas: \emph{i)} knowledge workers and well-being (\S\ref{subsec:rw_1}); and \emph{ii)} the role of VR in promoting well-being (\S\ref{subsec:rw_2}).

\subsection{Knowledge Work and Well-being}
\label{subsec:rw_1}
Knowledge workers face a number of challenges due to technological advances, hybrid work, and shifting organizational structures~\cite{constantinides2022future, rudnicka2020eworklife}. While these changes bring flexibility, they also introduce stressors (e.g., long hours, lack of meaningful work, and poor work relationships) which can lead to alienation~\cite{nair_exploration_2010}. Constant connectivity fragments attention with frequent interruptions~\cite{soto_observing_2021}, while open-plan offices and remote work exacerbate distractions, isolation, and blur work-life boundaries~\cite{teevan_new_2021}. Together, these factors contribute to stress, burnout, and mental health problems, which ultimately reduces engagement and work outcomes~\cite{who_burnout_2019}. Therefore, employee well-being is vital to organizational success.

Knowledge work requires environments that support autonomy and creativity~\cite{mladkova_knowledge_2011} because it involves processing complex information and making decisions without clear guidelines~\cite{septiandri2024potential}. However, organizational tools and technologies focused on standardization and efficiency can hinder these needs~\cite{karr-wisniewski_when_2010}. Tools such as task managers and AI-driven automation might reduce cognitive load~\cite{das2023focused} but could also perpetuate overwork and stress~\cite{leshed_i_2011, mark_effects_2018}. HCI research has recently increasingly moved from productivity-focused agendas to the holistic experience of technology use in work settings \cite{guillou_is_2020, kim_understanding_2019}, through well-being-focused solutions such as mindfulness apps or tools for supporting emotional resilience~\cite{howe_design_2022}. However, these technologies often face challenges in adoption as they are treated as optional add-ons rather than integral parts of work processes. Moreover, they frequently fail to provide workers with opportunities to detach fully from their workplace and tasks, limiting their ability to recharge and mentally recover.

\subsection{The Role of VR in Promoting Well-being}
\label{subsec:rw_2}
VR apps have been shown to support well-being through stress management~\cite{mazgelyte_effects_2021, pimentel_digital_2019}, emotional regulation~\cite{garcia-ballesteros_garden_2024, rodriguez_vr-based_2015}, and physical rehabilitation~\cite{camporesi_vr_2013}. Recent studies have explored their application in workplace well-being~\cite{Riches2023Virtual, Naylor2020A}, showing that VR interventions can reduce anxiety and negative mood states through nature-inspired or abstract environments and guided meditation~\cite{8252142, adhyaru2022virtual}. Thoondee and Oikonomou~\cite{8252142} demonstrated VR's stress-reducing effects for office workers, while Heyse et al.~\cite{heyse} enhanced relaxation by tailoring content to users’ emotions.

Nevertheless, many VR apps lack specificity for knowledge workers, often targeting general well-being instead of unique stressors and cognitive demands~\cite{8252142}. Hardware discomforts, such as prolonged headset use, limit adoption~\cite{kim_effect_2021}, and logistical issues in open-plan offices further complicate implementation. Privacy concerns surrounding data collection hinder trust~\cite{kaminska}, and sustaining engagement remains difficult, as the novelty of VR tools often wears off~\cite{Riches2023Virtual}. Despite these barriers, VR can provide distraction-free environments for relaxation and mindfulness~\cite{Naylor2019Augmented}. Features such as emotion-based adaptation~\cite{heyse} and biometric feedback~\cite{kaminska} can personalize experiences and boost engagement. Integrating VR within existing workplace tools such as calendars, can make tools more seamless and practical~\cite{chow_feeling_2024}. At the same time, transparent data practices are crucial for building trust~\cite{tahaei2023human}. By addressing these challenges, VR has the potential to become an effective and tailored solution for workplace well-being.
\smallskip

\noindent\textbf{Research Gaps.} 
%\todo{make sure this paragraph is linked with RQs and surfaces the complication that is stated in the abstract} 
Prior research on VR well-being often takes a solution-driven approach by evaluating predefined apps that target singular outcomes such as relaxation or focus \cite{adhyaru2022virtual}. However, this approach overlooks the multifaceted and fluctuating nature of workplace well-being, especially among knowledge workers whose needs can shift throughout the day. Moreover, many studies involve users only at the evaluation stage and exclude their perspectives from the design process \cite{Riches2023Virtual, heyse, 8252142}. As a result, existing tools may lack the flexibility and personal relevance needed for real-world uptake. In response, we adopted a human-centered approach integrating workers' experiences from early design to deployment.

\section{Author Positionality Statement}
We recognize that our positionality shaped the study's framing, design, and interpretation~\cite{havens2020situated}.  Our team, based at an Eastern European organization, includes one female and three male researchers with backgrounds in HCI, interaction design, AI, immersive and ubiquitous technologies.  Our positionality has shaped the study design in two ways: \emph{1)} prioritizing participants' lived experiences and perceptions over technology evaluation; and \emph{2)} balancing enthusiasm for immersive tech with critical reflection on feasibility. We also acknowledge that our institutional and personal perspectives informed our choices, while other interpretations may remain valid.

\section{Formative Study: Design Requirements}
\label{sec:design}
In Phase 1, we conducted a formative study involving interviews with knowledge workers to understand the challenges they face and their perceptions about using VR for workplace well-being. The study helped us identify design requirements for a VR app.
\smallskip

\noindent\textbf{Participants.} We recruited 10 knowledge workers (PD1-PD10) through professional networks. Participants included researchers and developers specializing in gaming, XR, and AI. They were aged 25-44 years, balanced across gender (5 male, 5 female), employed full-time, and varied in their work hours (6-10 hours daily). All participants regularly engaged in well-being activities (e.g., walking, meditation) and had prior VR experience. The study was approved by the ethics board of our institution. 
\smallskip

\noindent\textbf{Procedure.} Each participant completed a demographic survey prior to the semi-structured interviews. Interviews lasted 30-40 minutes and took place either online or in person, depending on participant preference. The interview protocol was informed by prior literature on workplace well-being and VR-based interventions, particularly work exploring strategies in digital health and relaxation-focused design in VR \cite{Riches2023Virtual, adhyaru2022virtual, Naylor2019Augmented}.
\smallskip

\noindent\textbf{Data Collection and Analysis.} 
Interviews were recorded, transcribed using \url{rev.com}, and reviewed for accuracy. We used thematic analysis for its flexibility in exploring user perspectives \cite{braun_using_2006}. We began with repeated reading and annotation to develop codes (e.g., \textit{coping mechanisms, interaction style: unstructured, personalization}). The codes were iteratively grouped into themes through discussions and synthesized into six design requirements (DRs).
\smallskip

\noindent\textbf{Design Requirements.} Participants described common well-being challenges at work, particularly physical inactivity and mental stress. Long hours at a desk often led to muscle soreness, headaches, and fatigue. PD3 noted, \textit{``I have to sit at the computer most of the time, and it’s kind of challenging,''} while PD5 shared, \textit{``I feel completely exhausted from work, and I don’t have the energy to do exercises.''} Poor ergonomics and the pressure to meet deadlines often made it difficult to take breaks or move around. Mental health concerns were also common, including stress, irritability, and difficulty relaxing after work. As PD6 put it, \textit{``Sometimes I get irritated, which affects my overall day, and I also need a lot of time to relax when I get home.''} These accounts point to the need for a flexible VR tool that can support a range of well-being needs, both physical and mental, informing our first design requirement:

\begin{itemize} \vspace{-0.2cm} \item[] \textbf{DR1: Support a range of well-being needs.} The tool should offer targeted support for physical (e.g., muscle tension, back pain) and mental (e.g., stress, overthinking) challenges through activities like guided breathing or stretching tailored to the user’s current state. \end{itemize}

\begin{figure*}[t!]
  \centering

  \includegraphics[width=.94\textwidth, height=4cm]{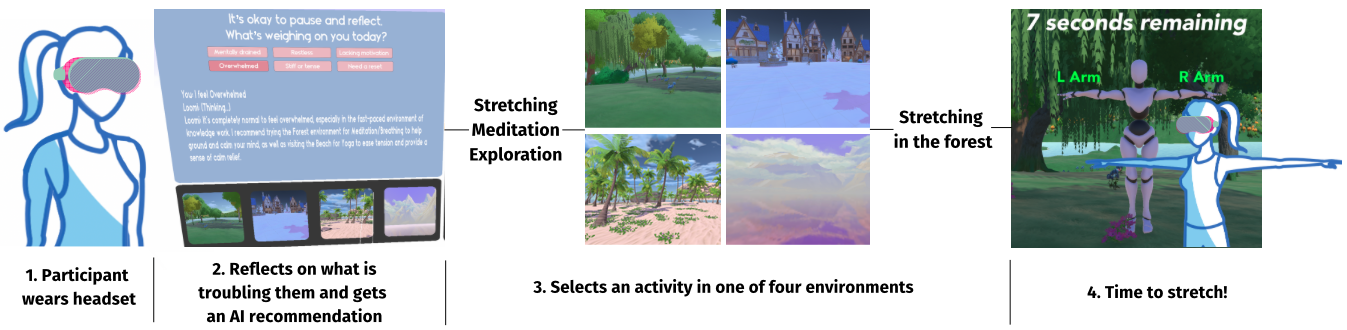} 
\caption {User Journey. (1) The user puts on the headset during work. (2) An AI agent, Loomi, greets them, asks what’s troubling them and suggests two activity-environment pairings. (3) Users can follow the suggestion or choose their own activity (stretching, meditation, or exploration) and environment (forest, snow, beach, or abstract). (4) Users are teleported to the chosen environment to begin the activity.}  
\label{fig:interaction}

\end{figure*}

\vspace{-0.2cm} When describing preferred VR activities, participants expressed interest in both structured, guided sessions and unstructured, self-directed experiences. Some favored structured approaches with features like progression levels (PD1, PD2), posture feedback through an avatar (PD4), or breath control guidance (PD3).  Others valued free-form or self-directed experiences to contrast the often structured nature of work (PD4, PD6). Participants envisioned environments they can explore in their own way. PD8 explained \textit{``you make yourself, get out of your comfort zone, you're not gonna say 'I need to do this', but say I'm gonna do whatever random thing I want in the moment.''} Ultimately, the ability to select between both structured and unstructured options was seen as a valuable feature from participants like PD7 \textit{``the option to either follow a guided session or just sit quietly [...] and look around.''} We captured this in our second design requirement: 

\begin{itemize}
\vspace{-0.2cm} \item[] \textbf{DR2: (Un)structured activity options.} Offer both guided (structured) sessions with instructions and free-form (unstructured) environments for self-directed activities. 
\end{itemize}

\vspace{-0.2cm} Participants were clear about how they wanted to use VR during the workday: short, focused sessions that fit between tasks. PD7 explained, \textit{``A quick 10-minute session between tasks would be ideal, enough to recharge but doesn’t disrupt my day.''} Rather than building VR into a fixed routine, most preferred using it as needed. PD6 said, \textit{``It should fit into my day without feeling like an extra task; it needs to flow with my routine.''} Several had negative experiences with well-being tools that required tracking or daily engagement, which quickly became tedious. PD1 preferred a more reactive approach: \textit{``When stress levels are up, I'd need breathing exercises. If I have pain in the back, I'd need [...] a light exercise.''} Participants also favored low-effort activities that did not require changing clothes or sweating, rather than physically demanding ones as proposed by previous literature \cite{haliburton_vr-hiking_2023}. As PD5 put it, \textit{``I feel exhausted from work, and I don’t have the energy to do exercises.''} Such preferences point to a need for quick and easy-to-start experiences that provide immediate relief without requiring sustained effort or habit-building, leading to our third design requirement: 

\begin{itemize}
    \vspace{-0.2cm} \item[] \textbf{DR3: Short and focused sessions with minimal disruption.} Ensure that sessions can be initiated quickly (e.g., 10-minute breaks) and provide immediate benefits without demanding excessive energy or commitment.
\end{itemize}

\vspace{-0.2cm} VR was seen as a way to mentally escape the office when physical breaks were not feasible. PD6 said, \textit{``If I could put on a headset and leave to a completely different environment for a few minutes, that would make a big difference.''} Preferred environments included natural settings like beaches (PD1, PD3), forests (PD4, PD6, PD8), and snow (PD4, PD7), as well as more abstract or minimal spaces (PD10). Participants emphasized choosing environments based on their mood or stress level. PD3 explained, \textit{``I would want something tailored to me, something that actually helps with my specific stressors, not just a generic relaxing landscape.''}  Stylized spaces were favored over hyper-realistic ones. PD9 noted, \textit{``Cartoonish, creative environments that let you explore or do unexpected activities would be much more engaging than just sitting at my desk.''} The appeal was not in how ``real'' a space looked but in how it felt and helped them disconnect from work, leading to two complementary design requirements:

\begin{itemize}
    \vspace{-0.2cm} \item[] \textbf{DR4: Diversity of environments.} Provide a wide selection of environments (e.g., natural, abstract) that users can choose from based on their current mood or context. Prioritize stylized aesthetics over realism to enhance emotional immersion.

    \vspace{-0.2cm} \item[] \textbf{DR5: Multipurpose VR environments.} Design environments that can support different types of activities (e.g., breathing or stretching), allowing users to decide what to do and where to do it based on their immediate needs.
\end{itemize}

\vspace{-0.2cm} Several participants emphasized the importance of personalization. Needs and stress levels varied, and participants wanted the tool to adapt accordingly (PD1, PD3, PD6, PD8, PD10). PD6 explained, \textit{``When I'm too stressed, I need someone to remind me how to calm down. More options to tell you what you could do would be good, depending on the situation.''} PD10 similarly suggested \textit{`` having personalized instructions for each user.''} For PD3, VR's strength was not in offering generic content, \textit{``like viewing a landscape,''} but in serving as \textit{``a problem solver to help me when I am going through things like panic attacks.''} Personalized suggestions were seen as key to making VR feel useful rather than generic.

\begin{itemize} \vspace{-0.2cm} \item[] \textbf{DR6: Personalized and context-aware recommendations.} Support users with suggestions that adapt to individual preferences, needs, and stress levels, offering relevant activities in response to their current situation. \end{itemize}
\begin{figure*}
  \centering
 \includegraphics[width=.23\textwidth, height=4.5cm]{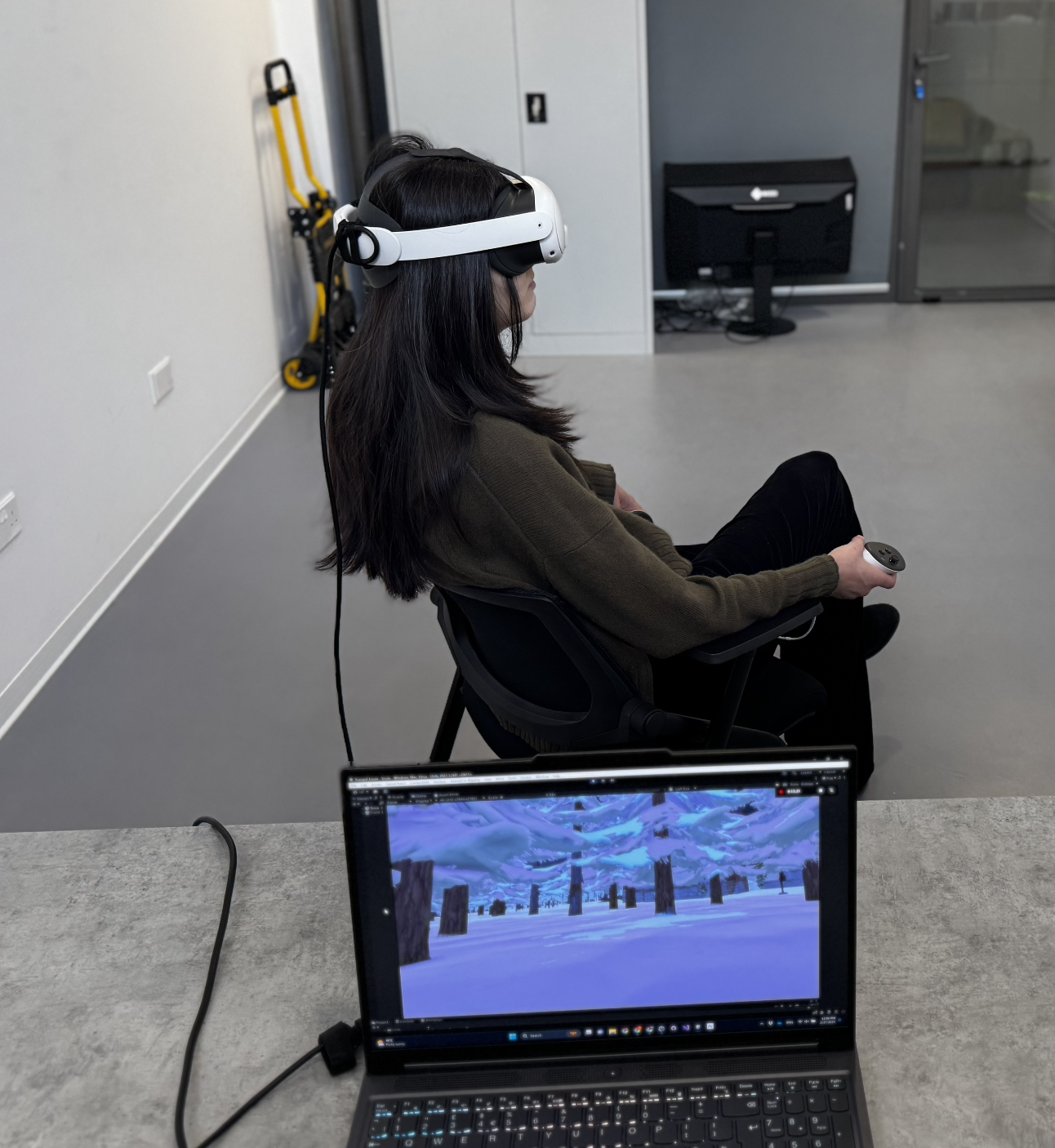} 
  \includegraphics[width=.2\textwidth, height=4.5cm]{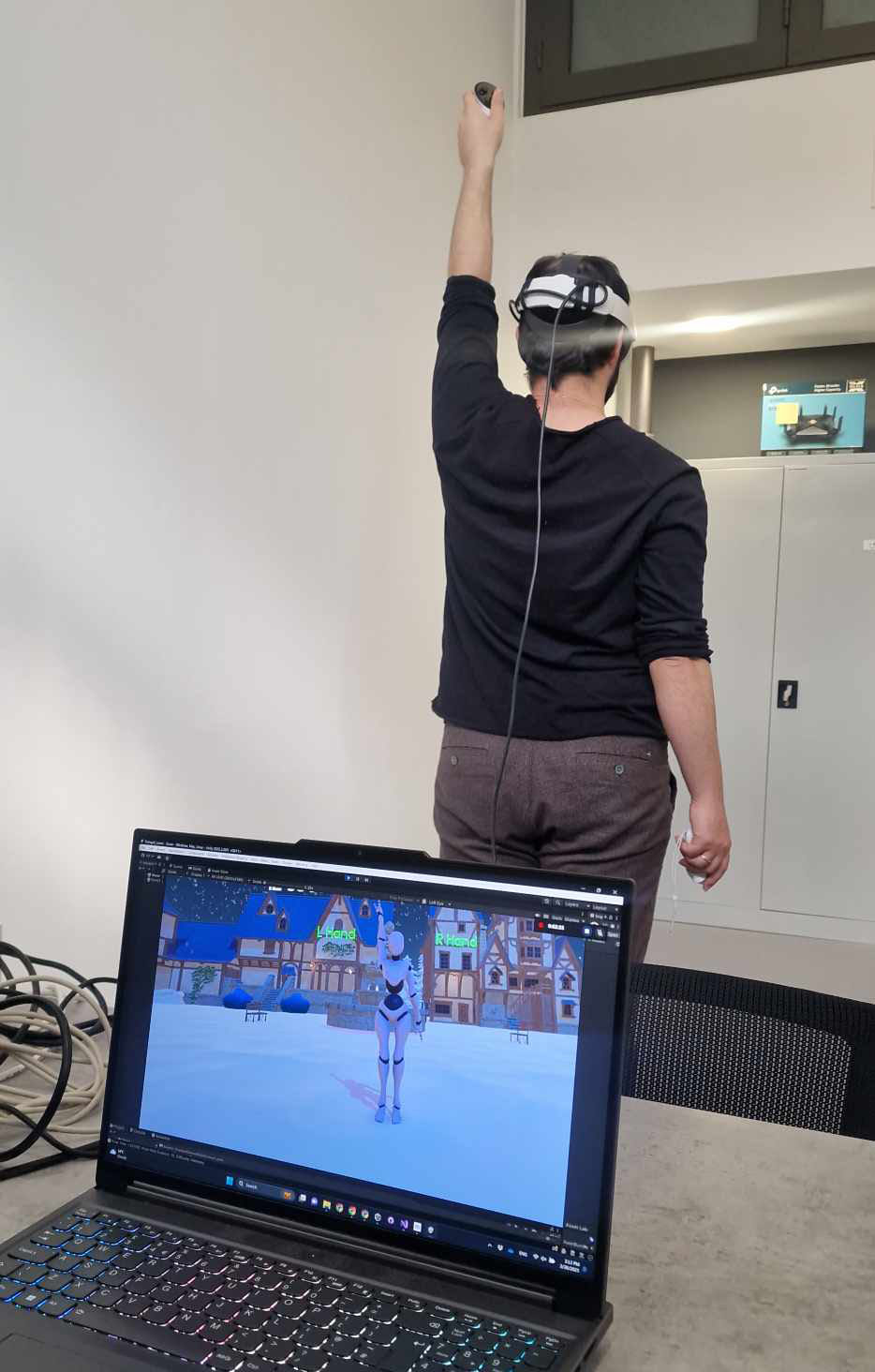} 
    \includegraphics[width=.18\textwidth, height=4.5cm]{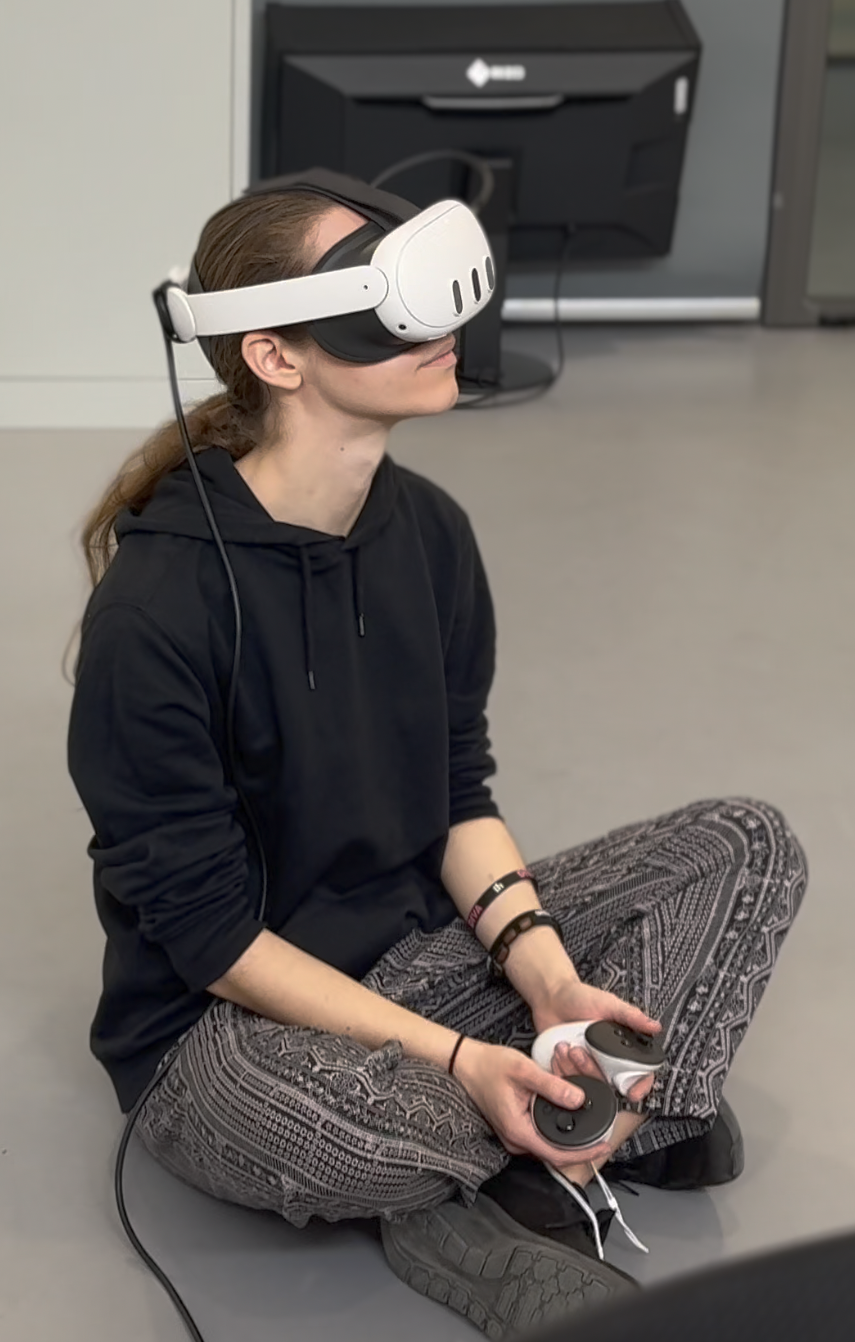} 
    \includegraphics[width=.28\textwidth, height=4.5cm]{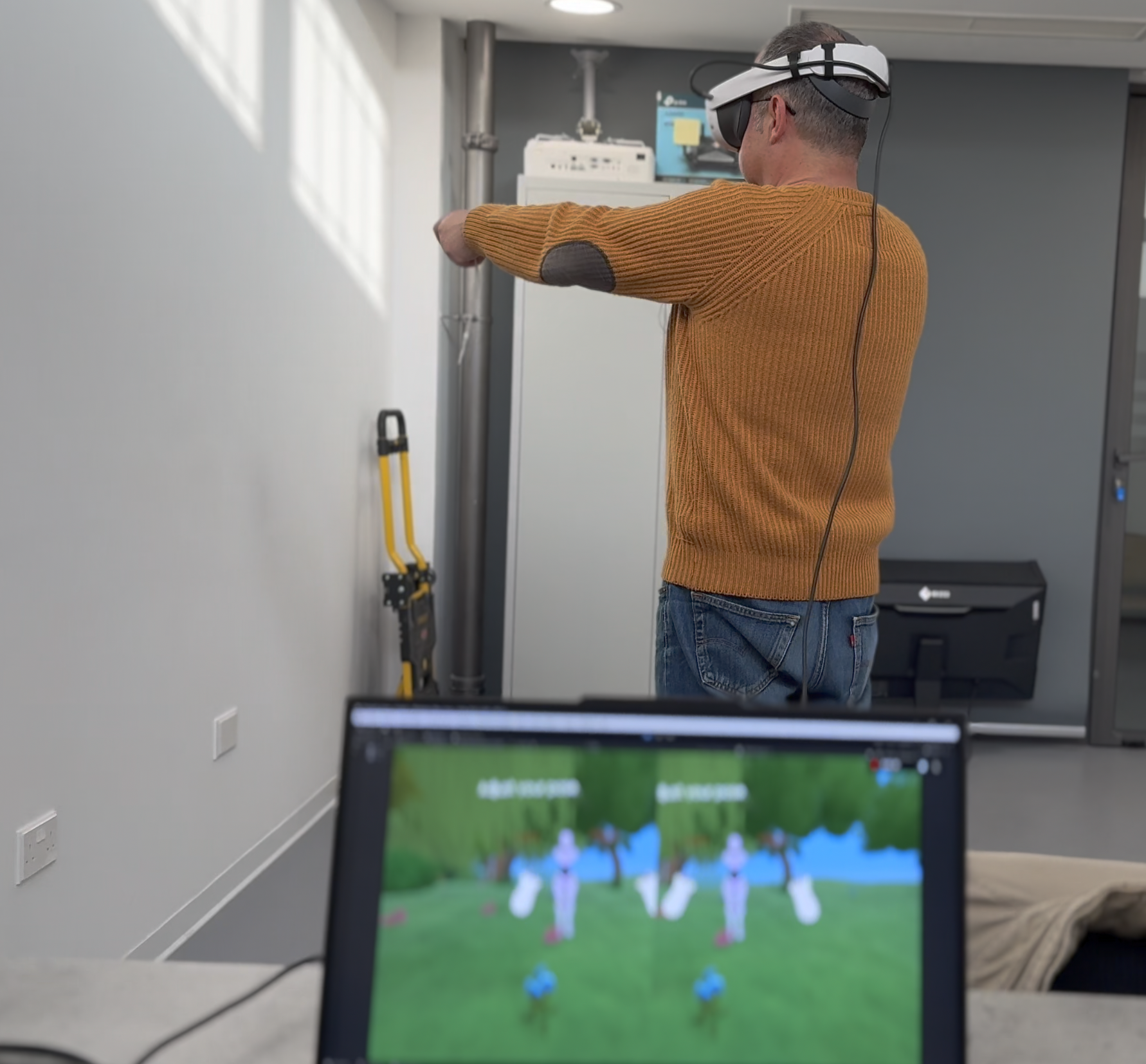}

\caption{Participants using Tranquil Loom during the workplace deployment. The in-situ deployment allowed us to observe how participants engaged in well-being activities in a realistic work context.}  
\label{fig:participants}

\end{figure*}

\section{Tranquil Loom}
\label{sec:app}

Based on the six design requirements identified in our formative study, we developed \textit{Tranquil Loom} (Figure \ref{fig:interaction}). It is a prototype VR well-being app designed to support short and restorative breaks during the workday. The app offers three types of well-being activities: stretching, guided meditation, and free exploration in four environments addressing DR1 (supporting a range of well-being needs) and DR2 (structured and unstructured activity options).

\smallskip
\noindent\textbf{Platform and Implementation.} Tranquil Loom was developed in Unity (version 6000.0.23f1) and deployed on the Meta Quest 3 using Meta Quest Link for PC streaming. Interaction was handled through Unity’s XR Interaction Toolkit and the Meta XR SDK.
\smallskip

\smallskip
\noindent\textbf{Environments.} The app includes four environments: forest, snowy landscape, beach, and an abstract space; these were assembled from Unity asset store models (Figure \ref{fig:teaser}). The environments had a stylized aesthetic as per the results of our formative study, and were selected based on participant preferences for both natural settings and more abstract and minimalist spaces. Each environment includes subtle animated elements (e.g., swaying trees, drifting snow). Users can perform any activity in any environment; this supports both our DR4 (diversity of environments) and DR5 (multipurpose use). We avoided overly realistic environments because creating highly detailed assets would have increased the risk of performance issues or motion sickness, while using 360 videos rather than 3D worlds would not have allowed for introducing the exploration mode and possibly reduce feelings of immersion \cite{ppali_keep_2022}.

\smallskip
\noindent\textbf{Sound Design.} All environments include spatialized nature sounds (e.g., water, wind, or bird songs). The abstract and home environments feature ambient music generated using \url{suno.ai}. The meditation activity includes a voice track (guided session) created using \url{elevenlabs.io}; this supports our DR1 (supporting a range of well-being needs) and DR3 (short, focused sessions). 

\smallskip
\noindent\textbf{User Journey and Personalization.} Upon launching the app, users begin in a calming home scene with two interactive panels. At the top of the first panel, a virtual agent named Loomi invites reflection with the prompt: ``What's weighing on you recently?''. Users can select from six preset feelings (i.e., mentally drained, restless, lacking motivation, overwhelmed, stiff or tense, need a reset), which were derived from our interview data and literature on workplace stress and emotional check-ins \cite{wagener_role_2021}. Based on the selected response, Loomi replies with a short message acknowledging the feeling and offers two suitable activity–environment suggestions. The interaction structure follows therapeutic communication strategies, that is, acknowledge, normalize, and recommend. These strategies are commonly used in mental health and counseling contexts \cite{cuff_empathy_2016}. Loomi's replies are generated using a large language model (LLM) via the OpenAI GPT-4o API, tailored through a prompt\footnote{\url{https://github.com/EX-MRG-CYENS-CoE/healthXR/}} that limits its behavior to a supportive and empathetic tone. While Loomi does not rely on open-ended natural language input, integrating an LLM enables lightweight personalization and context-sensitive recommendations in a conversational format. This feature supports DR6 (personalized and context-aware recommendations) and serves as a proof-of-concept for embedding LLM-driven agents within VR well-being apps~\cite{fang2024practicing}. We deliberately opted for multiple-choice input to minimize cognitive load and avoid typing-related frustration \cite{bhatia_text_2025}. However, this choice may have limited the depth of personalization we were able to introduce.
\smallskip

\noindent\textbf{Activities.} Tranquil Loom includes three activity modes: stretching, breathing, and open-ended exploration. To address DR4 (diversity of environments) and DR5 (multipurpose environments), the app allows users to choose one of three activities in any of four environments. Users can also end a session at any time and start a new one with a different activity or environment. This supports flexibility while keeping sessions short and focused as per DR3.
\smallskip

\noindent\emph{Stretching.} In the stretching mode, users follow a humanoid avatar demonstrating a sequence of six office-friendly stretches. These are primarily upper-body movements that avoid the need for floor-based poses or clothing changes, supporting DR3 (low-effort engagement). The app uses inverse kinematics to estimate the user’s body pose from head and hand positions. Above the avatar's hands, color-changing indicators (red to green) provide feedback on the correct form. Once alignment is achieved, a timer starts; if the user breaks form, it resets. The app includes three difficulty settings: bliss (10s), harmony (15s), and zen (30s). These were adapted from posture and movement guidance literature \cite{page_current_2012}. We chose not to include users' full-body representation. Instead, only the user's hands and guide avatar are visible. This choice allowed us to ensure the user's focus remained on mirroring the stretching poses.
\smallskip

\noindent\emph{Meditation.} In the meditation mode, the user's journey begins with a 2-minute voice-guided breathing session. The script is designed to support short, focused interventions during the workday, aligning with evidence that even brief meditation sessions can reduce stress and support cognitive clarity in knowledge workers \cite{zeidan_mindfulness_2010}.
\smallskip

\noindent\emph{Exploration.} In the exploration mode, users can freely navigate the environment using joystick movement. We considered implementing teleportation (common for reducing motion sickness), but ultimately chose joystick movement to support precise navigation and movement continuity \cite{buttussi_locomotion_2021}. To reduce discomfort, we limited movement speed and designed all environments as open spaces with minimal visual clutter. The ability to simply ``be'' in a space and explore it without a specific task supports DR2 (unstructured activity) and DR4 (emotional fit with the environment).
\smallskip

\section{User Evaluation}
\label{sec:evaluation}

\subsection{Methodology}
The evaluation study consisted of two phases: a familiarity session and a main study session. These were designed to introduce participants to the Tranquil Loom VR app and explore their experiences using it in a workplace setting.  \revision{The sessions were held in a quiet and spacious room to ensure minimal distractions, located in the city center to ensure accessibility to knowledge workers employed in nearby offices. We kept the familiarity session to 20 minutes and the main session to an hour so participants could join without interfering with their workday (even during lunch). If a participant needed a shorter session or more flexibility, we adjusted to fit their schedule.} We used a mixed-methods approach, collecting quantitative and qualitative data to understand participants' experiences with the VR app. Our institution's ethics board approved the study. 
\smallskip

\noindent\textbf{Procedure and Data Collection.} The first session was designed to familiarize participants with the Tranquil Loom app. This helped reduce potential barriers related to hardware discomfort or interface unfamiliarity before the main study and avoided any biases introduced due to the novelty effect. Upon arrival, participants signed informed consent and completed a demographics survey, which also included the Satisfaction with Life Scale (SwLS) questionnaire. A researcher introduced the VR headset and walked participants through how to use the app. Participants could ask questions and learn to use the experience in a low-pressure setting. 

Participants then booked a 1-hour timeslot within three days, during work hours, for the main study. At the start of the session, they completed two state well-being questionnaires: the State-Trait Anxiety Inventory (STAI-S) \cite{julian_measures_2011}, which measures current anxiety across 20 items (4-point scale), and the State Mindfulness Scale (SMS) \cite{ruimi_state_2022}, a 21-item tool (5-point scale) assessing present-moment awareness of bodily and mental states. Participants then used the app independently. Sessions were unstructured, with participants free to explore as they wished. Researchers gently concluded sessions exceeding 20 minutes to respect work schedules. Screen recordings captured navigation and feature usage, while researchers took observational notes on behavior and comments.

After the session, participants completed STAI-S and SMS again, along with the short form of the User Experience Questionnaire (UEQ-S) \cite{schrepp_applying_2014} and the User Engagement Scale (UES) \cite{obrien_development_2010}. UEQ-S measures usability and enjoyment via paired attributes on a 7-point scale (-3 to +3); UES evaluates engagement, satisfaction, and immersion across 30 items on a 5-point scale. Finally, participants took part in a 30-40 minute semi-structured interview. These explored their impressions of the app and broader reflections on the usefulness, limitations, and workplace role of VR for well-being. Participants were also asked about when and why they might use or avoid VR, how AI could be integrated meaningfully, and what features would help align such tools with their routines.
\smallskip

\noindent\textbf{Participants.}
We recruited 35 knowledge workers (PS1-PS35) via mailing lists and personal networks. These participants did not overlap with those in Phase 1. They were employed across a range of sectors, including academia, software development, engineering, design, finance, healthcare, administration, and human resources. All held primarily sedentary roles (5+ hours/day) and reported no history of VR-induced motion sickness. Ages ranged from 21 to 59.5 years ($\mu$ = 33.36, $\sigma$ = 9.45); 18 identified as women and 17 as men. Nationalities included British, Greek, Russian, Cypriot, Zimbabwean, Czech, Portuguese, and Canadian. In terms of education, 6 held a doctorate degree, 16 a master’s degree, 7 an undergraduate degree, and 2 had other qualifications. Participants described a range of short break activities during the workday, most commonly walking, socializing, or using digital devices. After work, they supported their well-being through personal strategies such as exercise, meditation, stretching, or occasional workouts.
Participants reported mixed experience with VR. The most common pattern was occasional use (n = 15), followed by never having used VR (n = 9), and yearly use (n = 5). A few used VR more frequently: monthly (n = 2), weekly (n = 1), or daily (n = 3). SwLS scores ranged from 10 to 26, with an average score of 20.91 ($\sigma$ = 3.27), indicating moderate overall satisfaction.
\smallskip

\noindent\textbf{Data Analysis.} For the quantitative analysis, we conducted descriptive and inferential statistical analyses to evaluate the effects of the app. First, we calculated summary statistics ($\mu$, $\sigma$, minimum, and maximum values) for our outcome variables, including the UEQ-S and the four subscales of UES (Attractiveness, Perspicuity, Efficiency, and Dependability). To examine pre-post differences, we conducted normality tests using the Shapiro-Wilk test on each pair of ``before'' and ``after'' measures, including STAI-S and SMS questionnaires. 

For the qualitative analysis, all interviews were audio-recorded and then transcribed with \url{rev.com}. Transcripts were then reviewed to make sure there were no discrepancies. The cleaned interview and observational data were qualitatively analyzed using thematic analysis~\cite{braun_using_2006}. We employed an iterative open-ended coding process, identifying data patterns related to our RQs. 
First, two researchers read the interview transcripts to re-familiarize themselves with the data. Then, during a discussion session, the research team reflected on the data and agreed on the key points for analysis. Following this, the interviews were split between the two researchers and coded using \url{Atlas.ti}, Code examples included: \textit{`non-goal oriented play', `customization', `familiarity over novelty', and ' visiting familiar places'}. Observational data from the notes during the VR sessions and from screen recordings were used to provide additional context.
 Over the course of the analysis stage, the two researchers had regular meetings to reflect on their codes and iteratively identify data patterns, gradually determining which patterns were most useful for becoming overarching themes, how they may be combined according to shared meanings, and which parts of the data could be discarded. Disagreements were resolved through successive rounds of synchronous review. In addition, discussions with the wider research team helped ensure that the final themes accurately captured the participants' experiences. Through the iterative process, we identified 4 overarching themes in relation to our RQs.
\smallskip

\subsection{Results}

\begin{figure}
    \centering
    \includegraphics[width=0.48\textwidth]{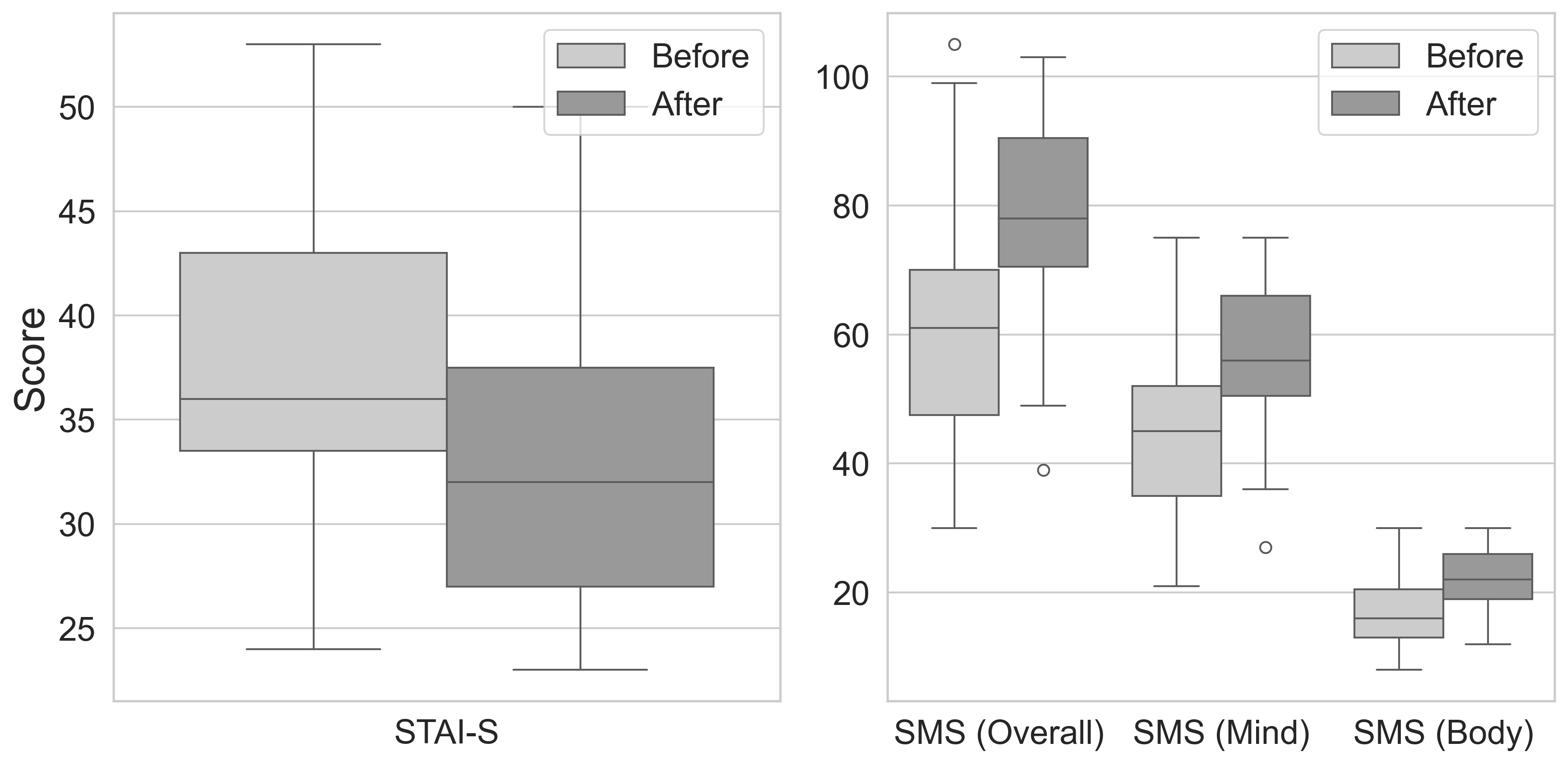}
    \caption{Boxplot comparisons of self-reported anxiety and mindfulness scores before and after using the app. The left panel shows a reduction in anxiety (STAI-S) from before to after the intervention. The right panel displays increases across all three dimensions of state mindfulness: overall awareness, mental engagement, and bodily presence (SMS subscales).}
    \label{fig:before-after-questionnaire}
\end{figure}
\subsubsection{Quantitative}
To assess the impact of the app on users' anxiety and mindfulness, we conducted paired-sample t-tests comparing scores before and after use; Shapiro-Wilk test showed normality across all our variables ($p$ $>$ .05). We found statistically significant improvements across all four outcomes (Figure~\ref{fig:before-after-questionnaire}). There was a significant reduction in anxiety levels as measured by $STAI-S$ ($t(34) = 4.88, p < .001$). Similarly, significant increases were observed in state mindfulness, with $SMS_{Overall}$ (t(34) = -6.70, $p < .001$), $SMS_{Mind}$ ($t(34) = -6.42, p < .001$), and $SMS_{Body}$ ($t(34) = -6.14, p < .001$).

To evaluate user perceptions of the app itself, we analyzed the responses to the UEQ-S and UES questionnaires. On the UEQ-S, the overall user experience was positive ($\mu$ = 1.73, $\sigma$ = 0.84). The UE scores further confirmed positive evaluations across dimensions: Focused Attention ($\mu$ = 3.89, $\sigma$ = 0.87), Perceived Usability ($\mu$ = 3.83, $\sigma$ = 0.81), Aesthetic Elements ($\mu$ = 3.96, $\sigma$ = 0.76), and Reward Factor ($\mu$ = 4.13, $\sigma$ = 0.54). 

\begin{table}[t!]
\centering
\footnotesize
\caption{Time participants spent in each scene ($\mu$ and $\sigma$ in seconds).}
\begin{tabular}{l c c}
\hline
\textbf{Scene} & \textbf{Mean Duration (s)} & \textbf{SD (s)} \\
\hline
Forest Stretching & 236.82 & 102.85 \\
Forest Exploration & 201.43 & 97.92 \\
Snow Exploration & 200.68 & 128.10 \\
Beach Stretching & 199.29 & 95.88 \\
Snow Stretching & 182.00 & 81.77 \\
Forest Meditation & 172.00 & 56.69 \\
Abstract Meditation & 143.23 & 72.36 \\
Beach Meditation & 116.50 & 47.32 \\
Abstract Exploration & 83.35 & 58.57 \\
\hline
\end{tabular}
\label{tab:scene_engagement}
\end{table}

To explore the effect of Loomi's LLM-powered recommendations on participants’ behavior within the VR experience, we analyzed scene-level interaction data. Each of the 12 available scenes (across four environments and three activity types) was paired with a binary indicator of whether it was accessed through an LLM suggestion. We excluded scenes where participants spent less than 30 seconds to focus on meaningful engagement. The results showed that participants spent, on average, 166.50 seconds in scenes they entered via an LLM recommendation, compared to 166.31 seconds in scenes they entered independently. This negligible difference suggests that while LLMs were used to guide scene selection, they had limited influence on how long participants remained engaged once in the scene. These findings indicate that momentary engagement may be more influenced by intrinsic scene qualities or personal preferences than by whether the scene was AI-recommended.

We also examined participants' overall engagement across the 12 available scenes, irrespective of whether the scene was recommended by the LLM (Table~\ref{tab:scene_engagement}). The most popular environment-activity combo was Forest Stretching ($\mu$ = 236.82s), while the least popular was Abstract Exploration ($\mu$ = 83.35s).

% To focus on meaningful engagement, we excluded visits where participants spent less than 30 seconds. The most time-intensive scenes were Forest Stretching ($\mu$ = 236.82s), Forest Exploration ($\mu$ = 201.43s), and Snow Exploration ($\mu$ = 200.68s), suggesting a strong interest in both movement-based and exploratory experiences. Among exercise-based scenes, Beach Stretching ($\mu$ = 199.29s) and Snow Stretching ($\mu$ = 182.00s) were also highly engaging. Interestingly, even some meditation scenes such as Forest Meditation ($\mu$ = 172.00s) and Abstract Meditation ($\mu$ = 143.23s) attracted sustained attention. In contrast, scenes such as Abstract Exploration ($\mu$ = 83.35s) and Beach Meditation ($\mu$ = 116.50s) were used for shorter durations. These results suggest participants were drawn to immersive and nature-themed environments, particularly those that supported active or contemplative engagement.

% Additionally, we explored the average times our participants spent in each activity independent to the scene (e.g., beach, forest, snow, abstract). We found that the most popular activity was stretching ($\mu = 192.43 s$), followed by exploration ($\mu = 165.2 s$), and meditation ($\mu = 138.92 s$).

% \todo{DONE. report result: average out all activities independent to the scene (e.g., beach stretching, forest stretching...) say which activity is the most popular} 
\smallskip

\subsubsection{Qualitative}

\noindent\textbf{VR as a ``drop-in'' Well-being Tool.} Our participants saw potential for VR at work as a drop-in well-being tool, to be used opportunistically rather than through scheduled sessions, supporting the findings of our formative study. 13 of our participants mentioned that for such a tool to be meaningful, it should be readily available without requiring extensive setup (PS3) or booking systems (PS9), catering to the unpredictable nature of work pressures. PS1 elaborated \textit{``I could see it being a drop-in thing. Booking presumes that you can plan for stresses at work, which most of the time doesn't really work that way. If my neck felt stiff after a long meeting, I might drop by and see if there's an available headset to use''.} PS24 and PS1 suggested that a dedicated space with enough headsets would support this kind of spontaneous use.

Moreover, our participants reiterated the need for diverse experiences that are suitable for short breaks between 5 and 20 minutes. The three activities we provided were seen as appropriate for this timeframe. Providing a means for quick stress relief or mental reset could help employees return to their tasks feeling more creative and focused. PS30 described \textit{``having this as a solution could even decrease the breaks because you take one break and it helps you relax and be more productive''.} Nevertheless, PS8 pointed out that for this to happen it is important for the experience to allow users to quickly detach from their surroundings: \textit{``the virtual environment should help you disconnect from the real world fast and be in a state that you forget where you are and you're fully engaged''.} PS12 and PS10, suggested usage limits were seen as a way to prevent overuse by employees. PS3 and PS10 explained that exiting the experience is as important; any type of time limit should be implemented in a way that takes the user smoothly out of the experience. Having a timer abruptly run out might undo the relaxation effect. 
\smallskip

\noindent \textbf{VR for Escaping the Office.}
Participants often described VR as a potential supplement or alternative, not a replacement, for real-life breaks or social interactions. PS5 explained \textit{``fundamentally it is there not to replace social breaks or any type of break, it is there as a thing that you can go for five minutes when you're feeling overworked''.} For VR to be considered a meaningful alternative, it had to offer something distinct from a typical 5 to10 minute off-screen break as PS16 pointed out: \textit{`` it should offer a different experience at what you do in your typical breaks at work, [...] an escape, an addition''.} Many valued VR for its ability to transport them away from the immediate work context. PS8 explained \textit{`` when you want to take a break and disconnect from everything it can help you. It gives you the chance to be in an environment that it's not easy to do in real world''.}  The included environments were found sufficient in creating a sense of disconnection from reality, which, according to PS5, could help \textit{``people who struggle to meditate with their eyes closed focus on the environment and in turn on the meditation''.}

Sound was consistently brought up as a crucial element in increasing the effectiveness of VR well-being experiences, with 12 participants pointing out that it was one of the main elements that helped them detach. Participants noted that realistic and soothing sounds, such as birds in a forest or the ocean at a beach, contributed significantly to the feeling of ``being there''. PS27 described that the audio \textit{``helped you get into the environment and live the present moment'',} and PS15 further noted that it helped them  \textit{``feel present in every scene''.}  Nevertheless, responses to sound were subjective. Wind in the snow scene, for example, was relaxing for some (PS14) and distracting for others (PS32). Similarly, the mellow music in the abstract scene was disliked by PS10, describing it as one of the main things they did not like about the experience, while for others like PS5, it was the main thing that kept them in that scene and helped them detach. PS5 spent an extensive amount of time in the abstract environment because they enjoyed combining the sound with the 'simplistic' environment. They explained \textit{``in everyday life the frequency of the change of the images that we have in front of us is so high. So, visually, I feel that I cannot find comfort in the environment. Whereas with the sound, I feel that the information is less, and I could focus on that''.} Lastly, participants also commented on the guided meditation. While Loomi's AI-generated voice was generally well-received, several wanted a slower pace.
\smallskip

\noindent\textbf{VR for Playful Well-being.} Activities that encouraged play and curiosity without causing stress were seen as a way of relieving stress and injecting fun into the workday. This allowed employees to mentally detach from work pressures (even momentarily) and \textit{``return back feeling refreshed and energized''.} as PS35 described. During the sessions, we observed our participants engaging in unplanned playful behaviors, often triggered by their curiosity to discover the environment. For example, participants tried to run on the water, knock on doors, or gather flowers. Such interactions provided a sense of fun and escape, according to PS32 and PS22. Even if they were not practically possible, the potential of their existence triggered participants' curiosity.  Exploration mode was particularly valued. PS26 described it as `fascinating', prompting them to \textit{``want to explore around and look at the entire landscape''}, while PS23 reflected that exploration \textit{`` taps into a primitive human need, wakes up the child inside us and their curiosity''} and helped them momentarily reconnect with themselves during a busy workday.

Although some participants like PS12 appreciated the passive nature of exploration as it allowed them to simply \textit{``to look around and absorb what is happening''}, others felt it lacked meaningful interaction besides moving (PS3). Suggestions to enrich the experience included non-goal-oriented interactions such as the ability to fly (PS3, PS4), going underwater (PS11), entering houses (PS12), picking up flowers (PS32), or looking for and finding animals (PS22, PS11, PS25, PS6). PS11 suggested \textit{``it would be nice to look for animals based on where the sounds are coming from. That would grab my attention and make me more focused''.} Easter eggs were also commonly requested as a way to make exploration more fun. As PS13 put it \textit{``I would have liked to explore more things, like finding Easter eggs or something different to spark my curiosity''.}

Beyond unstructured play, several participants suggested various game elements and play features that are more structured.  These included tasks like finding hidden items, collecting objects, or reaching locations (PS33, PS8). PS33 proposed \textit{``you could have points of reference, find that in the environment or try and look for or gather stuff, meaning in a forest environment you could collect flowers''.} Other suggestions included mini-games like fishing with scoring (PS20) or competitive challenges among employees, such as exercise-based leaderboards (PS3) and treasure hunts (PS33). Goals were seen as adding a sense of accomplishment and encouraging engagement. However, not all participants welcomed gamification. PS8 cautioned against overloading the experience with tasks that might feel like work or added pressure during a break. The key would be to find a balance where gamification enhances engagement without detracting from the primary goal of supporting well-being, as PS2 explained \textit{``It helps to have 5, 10 minutes of fun [...]. Maybe a fun mini-game. Not stressful one''}.

% \todo{Write about ai generated environments to keep things interesting. }

% \begin{itemize}
% \item Play whether structured or unstructured emerged during the interactions
% \item Participant asked for gamification
% \item but they also asked for non-goal oriented play. 
% \item Experiences that trigger curiosity / (e.g animal finding and so on and focus the attention)
% \end{itemize}
\smallskip
\noindent \textbf{AI for Personalizing the Experience.} \revision{The use of AI at the start of the app to prompt emotional reflection received mixed reactions. While some participants appreciated it as a gentle nudge, others found it unhelpful or overly simplistic.}

\revision{On the positive side, participants valued how the assistant helped reduce overthinking or offered helpful framing. \revision{PS21 used the AI to guide their experience, while PS14 shared: \textit{``It gives you an answer based on what you could do so you don't have to think [...] It directs you to the correct path''.}} Even participants who did not rely on the suggestions still found value in the option. \revision{As PS20 put it, \textit{``It [the AI] provides you guidance [...] but I don't know if it's necessarily helpful as I still went where I wanted to''.}}} \revision{Beyond the current implementation, many participants envisioned richer uses of AI. \revision{PS30 and PS25 imagined Loomi as a personal coach that adapts to their preferences over time. PS30 noted, \textit{``[AI could] learn from the way you use the app and guide you there''.} Others suggested tracking wellness progress (e.g.,  mood or performance). PS25 imagined Loomi offering summaries or posture feedback, while PS27 proposed real-time support: \textit{``If the app understands you're having difficulty, I can ask it, and the answer would come''.}} Participants also envisioned generative personalization, where the AI dynamically creates activities and environments. \revision{PS1 imagined conversations with Loomi leading to customized meditation or yoga based on how they felt that day. PS27 emphasized: \textit{``If I'm going to use it on an everyday basis, I would like different yoga options based on what I need''.} PS4 proposed scene variability: \textit{``It could generate the world [...] so it's different every time''.} PS33 added, \textit{``It could generate a cat doing funny things [...]. Something that cheers you up. Something that could be more like a mood booster''.}}}

\revision{On the negative side, several participants reported either not noticing the AI at all (PS8, PS32) or finding it too generic (PS22) or unnecessary (PS3). \revision{A key issue was the text-based interface. PS3 found it cognitively demanding during already overloaded workdays, proposing \textit{``Maybe a voice would have made me notice it more. The text didn't grab my attention''.}} Participants advocated for multimodal interactions (e.g., voice and embodied avatars) to make Loomi feel more engaging. \revision{PS20 suggested a playful character; PS23 imagined venting frustrations: \textit{`` I've been in front of my computer since 7:00 AM, so I don't want to see any more text. I want to talk to someone, want a voice.[...] If this is meant to help me relax, I think that a conversation could be helpful''.}}} \revision{While participants recognized the value of personalization, they also expressed strong reservations about the risks of integrating AI into well-being tools, particularly in workplace contexts. Trust emerged as a major concern. PS7 noted: \textit{``You don't know if the recommendation is good or not. So it's up to you to decide based on how much you trust it''.} PS8 proposed evaluating outcomes: \textit{``If it asks you whether you feel better after, it builds trust with AI tools''.} Participants also raised privacy concerns, especially if users began confiding in Loomi as they might with a therapist. \revision{PS1 warned: \textit{``If it gets to a point where people share intimate details, then the data becomes a big problem''.}} The workplace setting heightened these anxieties. \revision{PS3 shared: \textit{``If you want to complain about something at work and that is later used against you, it's a problem''.}} The possibility of data being used for targeted advertising or surveillance raised further ethical flags. Participants felt this tension between personalization and anonymity must be carefully managed. PS1 proposed a compromise: input data should be used in the moment but not stored (or reused) to train AI models.}

\section{Discussion}
\label{sec:discussion}

% Our work investigated how VR can support knowledge workers' well-being through short and self-directed breaks during work. Using a human-centred approach, we designed a VR well-being app called Tranquil Loom, and evaluated it with 35 knowledge workers. Quantitative data showed significant reductions in anxiety and increases in mindfulness following brief sessions. Stretching and open-ended exploration were especially engaging, which emphasize a need to balance structured and spontaneous experiences. Moreover, participants consistently emphasized the importance of sound design and environmental immersion, reinforcing that sensory qualities substantially shape VR's perceived effectiveness~\cite{adhyaru2022virtual}. 

We investigated how VR can support knowledge workers' well-being through short and self-directed breaks. Using a human-centered approach, we designed and evaluated Tranquil Loom with 35 participants. Brief VR sessions significantly reduced anxiety and increased mindfulness. Stretching and exploration were especially engaging, and highlight the need to balance structure and spontaneity. Participants also emphasized the importance of sound and immersive environments in shaping VR's effectiveness~\cite{adhyaru2022virtual}.

Additionally, our participants valued the idea of emotionally adaptive support despite usage data showed little difference between AI-selected and user-chosen activities. However, participants emphasized that AI should remain optional to preserve autonomy. Overall, these findings reinforce the value of flexible VR apps that adapt to users' momentary states without imposing expectations.

Drawing from our findings, we then discuss five design trade-offs for  VR workplace well-being apps:  assigned use \emph{vs.} spontaneous access; novelty \emph{vs.} familiarity; doing \emph{vs.} being; structure \emph{vs.} openness; and AI guidance \emph{vs.} user autonomy.

% Furthermore, our data uncovered five design trade-offs inherent in VR workplace well-being interventions: 

% Participants debated the benefits of \emph{novelty against the comfort of familiar environments}, considered \emph{active (doing) vs passive interaction (being)}, grappled with preferences for \emph{structured activities vs open-ended, playful exploration}, \revision{weighed assigned use \emph{vs.} spontaneous access, and reflected on the balance between AI guidance \emph{vs.} user autonomy}. Such tensions bring to the surface important considerations for future VR well-being designs, setting the stage for discussions on navigating these trade-offs to maximize user benefits effectively. 

\smallskip
% \noindent \textbf{Design trade-offs.}
% \smallskip
\noindent \textbf{\revision{Assigned Use \emph{vs.} Spontaneous Access.}} \revision{While structured well-being programs often rely on scheduled activities and check-ins, our findings suggest that such formalization may be counterproductive in VR. Participants preferred using the app as a drop-in tool and viewed it as a resource they could access informally and frequently, without pre-booking or planning. Short and situational uses between tasks (e.g., 5–20 minutes) were favored over fewer and longer breaks. Participants emphasized that when VR becomes another item on the to-do list, its restorative value is diminished.}

\revision{This highlights a tension between assigned use (which enables organizational oversight) and spontaneous access (which supports personal autonomy). Participants viewed booking systems, log-ins, and usage tracking as intrusive, likely to undermine trust and discourage use~\cite{bailey2006need}. Instead, they advocated for low-friction access through open spaces, minimal setup, and no monitoring. They framed well-being not as something to be scheduled or tracked, but as a right to be supported when needed.} Future implementations should explore ultra-accessible deployments such as quiet VR booths or side rooms directly within the office. Such setups could further lower barriers to use and promote spontaneous micro-breaks during the workday.

\smallskip

% \revision{This highlights a tension between assigned use (allows organizations to measure and manage participation) and spontaneous access (supports individual autonomy and emotional readiness). Booking systems, log-ins, usage quotas, and productivity-linked incentives were viewed as intrusive and undermining trust and discouraging use. These mechanisms clashed with the emotional privacy that participants sought and introduced friction that discouraged uptake~\cite{bailey2006need}. Instead, participants advocated for low-friction access: open spaces, minimal setup, and no monitoring. They framed well-being not as something to be scheduled or tracked, but as a right to be supported when needed. For managers and designers, this means shifting from measurement to trust, where emotional availability is prioritized over organizational accountability.}

\noindent \textbf{Novelty \emph{vs.} Familiarity.} Nature-based environments \cite{annerstedt_inducing_2013, browning_can_2020} and biophilic elements \cite{gattullo_biophilic_2022, garcia-ballesteros_garden_2024} are widely used in VR well-being tools, often associated with reduced stress and improved mood. Our participants similarly preferred natural scenes (i.e., forests and beaches) not for their realism, but for their emotional familiarity. The familiar settings were seen as easier to settle into and more effective for detachment during short breaks, echoing environmental psychology research that shows recognizable, low-effort environments support recovery in cognitively demanding contexts \cite{korpela_restorative_1996}. However, participants raised concerns about repetition. Even calming environments could feel stale without variation~\cite{ng_chenrezig_2023}, pointing to a tension between overstimulating dynamic scenes and static ones that become dull. Our stylized but recognizable environments appeared to strike a balance: emotionally grounding yet perceptually fresh.

The novelty effect \cite{kini_xr_2024} played a limited role. While novelty could attract initial interest, participants found too much of it would be distracting, especially in high-focus work contexts. They preferred intuitive and low-effort experiences that do not require adjusting to new interfaces or making additional choices. This aligns with prior research on the tension between stimulation and ease \cite{chirico_when_2019}, though few studies address how to maintain engagement without relying on novelty. For our participants, novelty worked best as subtle change. Playful surprises would be welcomed but not a reason for repeated use. Instead, returning to familiar environments was described as a way of emotional anchoring, with slight changes becoming a way to match changing moods or energy levels.

\smallskip

\noindent \textbf{Doing \emph{vs.} Being.} Our findings emphasize the need to balance active engagement (``doing'') with passive immersion (``being'') in VR well-being tools \cite{atherton_doing_2020}. Participants often explored their surroundings walking, attempting to pick up objects, or searching for hidden features, even when explicit interaction was not available. Such behaviors reflected spontaneous curiosity and a desire to momentarily step outside the structure of work. At other times, participants preferred stillness. We observed them sitting quietly, listening to the sounds, or focusing on visual details. These contrasting modes supported different forms of recovery: ``doing'' helped re-engage attention and counter boredom, while ``being'' offered a calming and low-demand experience that reduced stress.  Our observations resonate with established psychological frameworks, such as Kaplan’s Attention Restoration Theory \cite{kaplan_restorative_1995}, which emphasizes the restorative power of environments that balance gentle stimulation and effortless attention. The deliberate inclusion of both active and passive immersion in Tranquil Loom aligns with this theory, suggesting that successful interventions should address both. 

% Beyond this, several participants described moments of awe such as becoming absorbed in virtual spaces or feeling unexpectedly moved by sound or scenery. Awe, linked to perceptual vastness and emotional well-being~\cite{he_i_2024}, may have been enabled by Tranquil Loom's environments, especially during open exploration.
Beyond this, several participants described moments that suggested a sense of awe; becoming absorbed in vast virtual spaces, ``forgetting where they are'', or feeling unexpectedly moved by sound or scenery. Awe is often linked to perceptual vastness and a need to mentally accommodate the experience, and has been associated with reduced self-focus and increased emotional well-being \cite{he_i_2024}. While not explicitly designed for this, Tranquil Loom's stylized and expansive environments may have enabled such responses, particularly during open-ended exploration.
\smallskip

\noindent \textbf{Structure \emph{vs.} Openness.}
The tension between structure and openness highlights a core design challenge: how to support presence without prescribing behavior. Our findings suggest the value of VR well-being tools lies not only in the activities offered, but in allowing users to engage on their own terms. Participants moved fluidly between active and passive states, guided more by how they felt than by predefined tasks. This was especially evident in exploration mode, where users engaged in spontaneous, playful acts such as looking for animals, imagining hidden features, or trying to fly. These unstructured behaviors were described as refreshing and emotionally satisfying. Rather than distractions, they were expressions of well-being where goal-driven thinking is suspended, therefore allowing alternative forms of attention\cite{atherton_doing_2020}. Although play is rarely foregrounded in workplace well-being tools, our findings align with research showing that light, curiosity-driven engagement can restore attention and support emotional regulation \cite{kaplan_restorative_1995}. Participants proposed low-pressure interactive features, like Easter eggs \cite{lakier_more_2022}, as gentle prompts for discovery and delight, offering emotional connection without cognitive strain. %
\smallskip

%Together, these insights suggest that future VR well-being tools should not frame structure and unstructured activity as a binary, but consider how to support fluid movement between the two. Supporting curiosity and play is as important as offering relaxation and focus.
\smallskip
\noindent\textbf{\revision{Guidance \emph{vs.} Autonomy.}} \revision{Participants were divided on whether AI suggestions added value to their experience or interfered with their own agency. Some appreciated Loomi's ability to reduce decision fatigue, especially when feeling overwhelmed or mentally drained. The assistant was seen as a gentle prompt that helped them get started without overthinking (i.e., guiding without prescribing).}
% ~\cite{luger2016like}
\revision{Yet many others preferred to ignore Loomi's suggestions entirely, choosing instead to follow their instincts or needs in the moment. Even those who acknowledged the AI's input often reframed it as a ``nice-to-have'' rather than something they actively relied on. Participants emphasized that AI should not override user agency, but offer lightweight scaffolding, perhaps suggesting ``a good place to start'' or adapting over time as preferences evolve. The tension here is between offering guidance and supporting self-direction. However, this reflects a broader trade-off: should the tool guide users toward pre-determined ``helpful'' paths, or should it simply open space for reflection and allow the user to lead? Too much guidance risks feeling prescriptive~\cite{binns2018s}, while too little may leave users feeling unsupported, especially during vulnerable moments.}

% Throughout our research, we found that non-goal-oriented play and playful interactions significantly enhanced the value of VR as a drop-in well-being tool in workplace settings. Unlike traditional gamification, which tends to focus on structured tasks or clear objectives \cite{liarokapis2022improving}, unstructured interactions, such as open exploration, helped participants mentally disengage from work and find immediate relief. These interactions tapped into natural curiosity, a quality widely recognized in the literature as important for stress reduction and creativity \cite{atherton_doing_2020}. In our study, moments like running across virtual water or trying to interact with flowers allowed users to reconnect with a playful mindset. At the same time, the study revealed a balance must be struck between play and overstimulation. Participants suggested small and low-effort enhancements  to increase engagement without adding cognitive load. This aligns with Riches et al. \cite{Riches2023Virtual}, who emphasize the value of curiosity-driven interactivity in VR well-being tools. 

\section{Design Implications}
\label{sec:design-implications}
\revision{Drawing from the design trade-offs, we discuss key considerations for developing effective VR well-being apps for the workplace. Together, these implications call for a (re)thinking of how VR well-being apps are positioned and designed: \emph{not as solutions to optimize workers, but as emotionally intelligent tools that support autonomy, trust, and self-directed recovery}.}
% \revision{Drawing from the design trade-offs, we discuss key considerations for developing effective VR well-being apps for the workplace. These implications call for (re)thinking how VR well-being apps are positioned and designed: \emph{not to optimize workers, but to support autonomy, trust, and self-directed recovery}.}
\smallskip

\noindent\textbf{\revision{Enable spontaneous use through low-friction access.}} \revision{Participants preferred short (5-20 minutes) situational use over scheduled sessions. VR well-being apps should be easily accessible without pre-planning or tracking through dedicated spaces, minimal setup, and optional use models. Well-being should not be another task to manage, but a resource to reach for when needed.}
\smallskip

\noindent\textbf{\revision{Design for emotional continuity, not novelty.}} \revision{Our findings challenge the assumption that novelty is necessary for sustained engagement. Instead, emotionally familiar environments supported relaxation and emotional anchoring. Rather than offering entirely new experiences, designers should support subtle evolution over time (e.g., shifting light, ambient sounds) that preserves the emotional tone and, where possible, create feelings of `awe'.  The `novelty effect', when used, should support the user's sense of ease.}
\smallskip

\noindent\textbf{\revision{Support fluid transitions between modes of engagement.}}
\revision{Well-being is not a fixed state, and users' needs may shift even within a single session. Participants naturally moved between active and passive engagement based on how they felt. VR apps support different modes of presence. Designers should enable transitions between ``doing'' and ``being'' to accommodate fluctuating attention and energy levels. Passive features such as ambient audio, as well as optional play elements (e.g., hidden objects or responsive scenery) may offer restorative engagement without pressure.}
\smallskip

\noindent\textbf{\revision{Embed structure without enforcing it.}}
\revision{Our findings suggest that future VR well-being tools should not frame structure and unstructured activity as a binary, but consider how to support fluid movement between the two. Supporting curiosity and play is as important as offering relaxation and focus. While structured activities can offer purpose, they should be offered as optional scaffolding available when needed, but never imposed. Design should empower users to dip in and out, choose their own paths, and interpret well-being on their own terms. Even ``guidance'' features should default to low-commitment interactions, with space for or free-form use.}
\smallskip

\noindent\textbf{\revision{Design AI guidance as optional and contextual scaffolding.}} \revision{Participants valued AI support when it was framed as gentle and optional guidance that reduced decision fatigue during moments of stress. However, they strongly resisted prescriptive or overly directive suggestions. Designers should treat AI as a companion that offers context-aware prompts without assuming authority. This means enabling users to easily ignore, override, or disable suggestions, while still allowing for personalisation if and when users seek it. Adaptive tools should learn preferences subtly and transparently to support autonomy rather than structuring behavior.} Additionally, future VR well-being apps could leverage generative AI to allow users to co-create or modify their environments (e.g., turning a beach into a bay or adding unique features), which potentially may foster a stronger sense of ownership and encourage return visits.

\section{Limitations and Future Work}
\revision{In Phase 1, we acknowledge two limitations. First, the sample size was small and included participants recruited through professional networks with prior experience in XR, gaming, or AI. Second, the tech-savvy sample may have been more receptive to VR, thus limiting the generalizability of the design requirements. Future work should include participants with varying levels of technological familiarity to understand broader needs and adoption barriers.}

% \revision{In Phase 2, we acknowledge five major limitations.
% First, the deployment was short-term, limited to one-off sessions in a specific workspace setting, and the participant sample was tech-savvy, which might not reflect broader populations.  Second, the tethered VR setup reduced portability compared to standalone headsets. Third, the app supported only low-effort movement and limited input styles, which might have affected appeal, and supported one activity per session, which might have disrupted the flow.
% Fourth, we prioritized practical application over technical novelty when designing the app. This was an intentional choice as the focus of our work was on exploring how existing immersive tools can be used in practical ways to support well-being at work.  Finally, the AI assistant provided static suggestions that might have limited personalization. Future studies should explore longer-term use with adaptive features, test the app in diverse work settings, and support less tech-confident users with smoother activity transitions.}

\revision{In Phase 2, we acknowledge five limitations. First, the short-term deployment in a specific workspace with relatively tech-savvy participants may limit generalizability. Second, the tethered VR setup reduced portability versus standalone headsets. Third, the app only supported low-effort movement, limited input, and one activity per session, which may have affected engagement. Fourth, we prioritized practical application over technical novelty, focusing on the use of existing immersive tools for workplace well-being. Finally, the AI assistant provided static suggestions, limiting personalization. Future work should explore longer-term, adaptive use in diverse work settings and support for less tech-confident users.}

% Third, while the app supported low-effort physical activity, it lacked full-body movement or varied input modes, which may limit appeal. Fourth, users could access only one activity per session (requiring menu navigation to switch), which may have disrupted the flow. Fifth, participants were largely tech-savvy, which may not generalize to broader user groups. Sixth, the AI assistant offered static suggestions without learning over time, thus limiting personalization. Finally, the system did not introduce new technological features or interaction techniques. This was an intentional choice: the focus of our work was not on technological novelty, but on exploring how existing immersive tools can be used in practical ways to support well-being at work. Future studies should explore longer-term deployments to understand sustained use, particularly with adaptive or context-aware systems. Testing in varied work settings with standalone devices would offer more realistic insights into everyday integration. Including less tech-confident users and enabling smoother transitions between activity types could help create more inclusive and flexible VR well-being tools.}

\section{Conclusion}
\label{sec:conclusion}
% Our work explored how VR can support workplace well-being through Tranquil Loom, a VR app designed for short, self-directed breaks. Through a two-phase mixed-methods study, we found that even brief VR sessions can positively impact mindfulness and anxiety while offering users a surprising sense of agency and playfulness, particularly through open-ended exploration. Our participants embraced VR as a spontaneous tool to navigate fluctuating states of well-being during the workday. Our  findings revealed design tensions between AI-driven personalization and autonomy, as well as structure and openness. Well-being in VR is not only about what users do but how freely
% they are allowed to do it. Future work should explore how adaptive VR experiences might fluidly respond to changing user contexts without overriding user intent. 

We explored how VR can support workplace well-being through Tranquil Loom, which is a VR app for short and self-directed breaks. Even brief VR use improved mindfulness, reduced anxiety, and promoted agency and playfulness. However, design tensions emerged between personalization and autonomy, and between structure and openness. Future work should explore adaptive VR that responds to users' changing needs without overriding intent.

% \clearpage
% \balance

\acknowledgments{
The work has received funding from the European Union’s Horizon 2020 Research and Innovation Programme Grant Agreement No. 739578 and the Government of the Republic of Cyprus through the Deputy Ministry of Research, Innovation and Digital Policy. It has also received funding from the European Union under grant agreement No. 101093159.}
% \nocite{*}
%\bibliographystyle{abbrv}
\bibliographystyle{abbrv-doi}

\bibliography{main}
\end{document}